
\input phyzzx.tex
\input phrdefs.tex
\def\ibid{{\it ibid.}}
\input tables.tex
\def\rbtau{R_{b/\tau}}
\def\rbtaup{R_{b'/\tau'}}
\def\lamt{\lambda_t}
\def\lamb{\lambda_b}
\def\lamtau{\lambda_{\tau}}

\def\lamtp{\lambda_{t'}}
\def\lambp{\lambda_{b'}}
\def\lamtaup{\lambda_{\tau'}}
\def\lamnup{\lambda_{\nu'}}
\def\mtp{m_{t'}}
\def\mbp{m_{b'}}
\def\mnup{m_{\nu'}}
\def\mtaup{m_{\tau'}}

\def\prdj#1{{\it Phys. Rev.} {\bf D{#1}}}
\def\npbj#1{{\it Nucl. Phys.} {\bf B{#1}}}
\def\prlj#1{{\it Phys. Rev. Lett.} {\bf {#1}}}
\def\plbj#1{{\it Phys. Lett.} {\bf B{#1}}}

\def\ptpj#1{{\it Prog. Theor. Phys.} {\bf {#1}}}

\def\anti{\overline}
\def\pbi{~{\rm pb}^{-1}}

\catcode`\@=11 

\def\t1{{\tilde 1}}

\def\gev{\,{\rm GeV}}
\def\tev{\,{\rm TeV}}

\def\rta{\rightarrow}

\def\tanb{\tan\beta}
\def\mt{m_t}
\def\mb{m_b}
\def\mz{m_Z}

\date{June, 1994}
\Pubnum{$\caps UCD-94-25$\cr}
\titlepage
\baselineskip 0pt
\bigskip
\centerline{\bf GAUGE-COUPLING UNIFICATION AND THE MINIMAL SUSY MODEL:}
\vskip .05in
\centerline{\bf A FOURTH GENERATION BELOW THE TOP?}
\smallskip
\centerline{J.~F.~GUNION$^{(a)}$, Douglas W. McKAY$^{(a),(b)}$ and H.
POIS$^{(a)}$}
\smallskip
\centerline{(a) \it Davis Institute for High Energy Physics,}
\centerline{\it Department of Physics, U. C. Davis, Davis, CA 95616}
\centerline{(b) \it University of Kansas, Department of Physics and
Astronomy,}
\centerline{ \it Lawrence, Kansas, 66045\foot{Permanent Address.}}
\vskip .1in
\centerline{\bf Abstract}
We explore the possibility of a fourth generation in
the gauge-coupling-unified, minimal supersymmetric (MSSM) framework.
We find that a sequential fourth generation (with a heavy neutrino $\nu'$)
can still fit, surviving all present
experimental constraints, provided $\lambda_b(M_U)=\lambda_\tau(M_U)$
Yukawa unification is relaxed.
For the theory to remain perturbative up to $M_U$,
the new leptonic generation must
lie within reach of LEP-II and the new $b',t'$ must have masses within
the reach of the Tevatron. For example, for $\mt>150\gev$ we find
$m_{\nu'},m_{\tau'}< 86\gev$, $\mtp<178$, and $\mbp<156\gev$.
Experiments at Fermilab are already sensitive to the latter
mass regions; we comment on direct $b'$ searches and
on the $m_{t'}\simeq \mt$ case in light of new CDF data.
Discovery may involve novel decay signatures;
however, CDF and LEP-II will confirm or exclude
an MSSM fourth generation in the near future.

\bigskip
\leftline{\bf 1. Introduction}
\smallskip

The minimal supersymmetric model (MSSM) prediction of gauge-coupling
unification consistent with experiment
\Ref\unif{P. Langacker, in Proceedings of the
PASCOS-90 Symposium, eds. P. Nath and S. Reucroft (World Scientific, 1990);
P. Langacker and M. -X. Luo, \prdj{44}, 817 (1991); U. Amaldi, W. de Boer, and
H. Furstenau, \plbj{260}, 447 (1991); J. Ellis, S. Kelley and D. V. Nanopoulos,
{\it ibid.} {\bf 260}, 131 (1991).}\
makes it the most successful approach for relating physics at the unification
scale, $M_U$, to physics below a $\tev$, and has led to many detailed
investigations of Yukawa and superparticle phenomenology within the MSSM
context.
\REF\DHR{S. Dimopoulos, L. Hall
and S. Raby, \prlj{68}, 1984 (1992); \prdj{45}, 4192 (1992); P. Ramond, R.G.
Roberts and G. Ross, \npbj{406}, 19 (1993).}
\REF\BARGERetal{V. Barger, M. S. Berger, T. Han and M. Zralek,
\prlj{68}, 3394 (1992); H. Arason, D. J. Castano, E. J. Piard and P. Ramond,
\prdj{47}, 232 (1993); V. Barger, M. S. Berger and P. Ohmann, \prdj{47}, 1093
(1993); C. Albright and S. Nandi, Fermilab-PUB-93/316-T,
OSU-Preprint 282; G. L. Kane, C. Kolda, L. Roszkowski and J. Wells,
UM-TH-93-24, Oct. 1993.}
\refmark{\DHR,\BARGERetal}
In this paper we return to an old question that has not been
re-examined using the most recent experimental constraints
and theoretical inputs.
Namely: how much room (if any) is there to add a fourth
generation to the ``MSSM-plus-unification" framework?\foot{We identify
the fourth generation by the CKM matrix hierarchy: $V_{tb}\simeq V_{t'b'}>
V_{t'b}\sim V_{tb'}$.}  Indeed, aside from
occasional studies\Ref\AGRAWALetal{P. Agrawal and W. -S. Hou,
\prdj{46}, 1022 (1992); P. Agrawal, S. Ellis and W. -S. Hou, \plbj{256},
289 (1991); S. King, {\it ibid},{\bf 281}, 295 (1992); C. Hill and E. Paschos,
{\it ibid.}, {\bf 241}, 96 (1990); W.-S. Hou and R. G. Stuart, \npbj{349},
91 (1991); M .Sher and Y. Yuan, \plbj{285}, 336 (1992); H. Fritzsch,
\plbj{289}, 92 (1992); B. Mukhopadhyaya and D.P. Roy, \prdj{48}, 2105
(1993).}\
and experimental searches, broad interest in a possible fourth
generation ended abruptly when Mark II and detectors at LEP
found that only three light neutrinos are allowed.
\Ref\Nneutrino{Mark II Collaboration, G. S. Abrams et al.,
\prlj{63}, 2173 (1989); L3 Collaboration, B. Advera et al., \plbj{231}, 509
(1989); OPAL Collaboration, I. Decamp et al, {\ibid}, {\bf 231}, 519
(1989); DELPHI Collaboration, M. Z. Akrawy et al. {\ibid}, {\bf 231},
539 (1989).}\ To
provide non-zero mass for the fourth generation $\nu'$ in the minimal
manner, we introduce an $SU(3)\times SU(2)\times U(1)$ gauge singlet,
right-handed neutrino $\nu_R$\Ref\KING{S. King, Ref.~[\AGRAWALetal].}\
\foot{We note that in the minimal $SU(5)$ model, adding a
$\nu_R$ would lead to a Dirac mass that would naturally be of the
order of the quarks or leptons due to the $\bf 5_f \bar 5_H \nu_R$ coupling.}
and its corresponding superfield.

Previous works have argued against the
inclusion of a fourth generation in the MSSM context
on the basis of rather specific assumptions
and/or approximations. For instance, if it is assumed that
the $\tau$ and $b$ Yukawa couplings unify, \ie\
$R_{b/\tau}\equiv \lambda_b(M_U)/\lambda_\tau(M_U)=1$
(as motivated primarily by the minimal
$SU(5)$ GUT theory\foot{By `minimal' we mean the standard $SU(5)$
${\bf 5,\bar 5,24}$ Higgs representation content.}),
inclusion of a fourth generation in the Standard Model (SM) or the MSSM
spoils the good
(given the uncertainty in $m_b$\Ref\CHANOWetal{M. Chanowitz, J. Ellis and
M. Gaillard, \npbj{128}, 506 (1977); A. Buras, J. Ellis, M. Gaillard and
D. V. Nanopoulos, {\it ibid.}, {\bf 135}, 66 (1978); M.B. Einhorn and D.R.T.
Jones, {\it ibid.}, {\bf 196}, 475 (1982);
J. Ellis, D. V. Nanopoulos and S. Rudaz,
{\it ibid.}, {\bf 202}, 43 (1992); a recent analysis of both SM and the MSSM
is given in H. Arason, D. J. Castano,
E.J. Piard and P. Ramond, \prdj{47}, 232 (1990).})
prediction for $m_b/m_\tau$\Ref\Nanojones{D. V. Nanopoulos and D. Ross,
\plbj{118}, 99 (1982); J. Bjorkman and D. Jones, \npbj{259}, 533 (1985).}\
obtained in the absence of a fourth generation.
Other early studies excluding a fourth generation employed
specific parameter choices
(often in conjunction with imposing exact Yukawa unification)
that limited $\mt$ to low values ($20\gev <\mt<120\gev$).
For example, Bjorkman and Jones\refmark\Nanojones\ assumed $\tan\beta=1$,
whereas larger values of $\tan\beta$ must be chosen
to allow for a heavier top-quark. We find that
if $\tan\beta>1$ is allowed and $R_{b/\tau}=1$ is relaxed (as motivated below),
then there is still significant room for a fourth
generation\foot{More than four generations are clearly excluded.
$N_g>4$ in the MSSM results in loss of asymptotic freedom for $SU(3)$ color,
and inability to evolve perturbatively to a point of gauge-coupling
unification; see Bjorkman and Jones, Ref.~[\Nanojones].}
of quarks and leptons (with masses below $\mt$), and their superpartners.

Relaxation of $\lambda_b(M_U)=\lambda_{\tau}(M_U)$ is motivated by
recent theoretical work indicating that
Yukawa unification at $M_U$ is not
necessarily expected given possibly large threshold effects at
SUSY breaking and ${\cal O}(M_U)$ scales,\Ref\threshold{P. Langacker and
N. Polonsky, \prdj{47}, 4028 (1993); {\ibid}, {\bf 49}, 1454 (1994);
V. Barger, M. Berger, P. Ohmann, and R. J. N. Phillips,
\plbj{314}, 351 (1993); G. Kane et al. Ref. [3]; G. Ross and R.G. Roberts,
\npbj{377}, 571 (1992); B.D. Wright, MAD/PH/812. For string Yukawa
threshold effects, see I. Antoniadis, E. Gava, K.S. Narain and T.R. Taylor,
\npbj{407}, 706 (1993).}\ and/or because there
may be no symmetry (global and/or local) in
the model at $M_U$ which leads to an exact equality among the Yukawa couplings;
this latter is the case for several candidate supergravity theories
(e.g. flipped $SU(5)$\Ref\JORGE{In
J. L. Lopez, D. V. Nanopoulos and A. Zichichi,
CERN-TH-713 8/94, $0.7<\lambda_b(M_U)/\lambda_\tau(M_U)<1$
is studied at the string scale.}) and for mild extensions of
the minimal $SU(5)$ model.\foot{For example, $R_{b/\tau}\not=1$ in general
when a $\bf 45$ Higgs representation
\Ref\Frampton{P.H. Frampton, S. Nandi and J.G. Scanio,
\plbj{85}, 225 (1979).} is added.}

The shift in focus to large top-quark masses, aside from requiring
that $\tanb>1$ be considered, has
led many authors to promote the naturalness of
infrared `fixed-point' solutions to the RG
equations,\Ref\fixedpoint{B. Pendelton and G. Ross,
\plbj{98}, 291 (1981); C. Hill, \prdj{24}, 691 (1981);
W. Bardeen, M. Carena, S. Pokorski and C. Wagner,
\plbj{320}, 110 (1994); J. Bagger, S. Dimopoulos and E. Masso,
\prlj{55}, 920 (1985).}\ for which a
large range of top Yukawa-coupling
($\lambda_t$) values evolves down to a narrow range of large values
at the weak scale; a close correlation between $\tanb$
and $\mt$ arises. We will discuss
whether or not such approximate `fixed-point' behavior
for some or all of the quark Yukawas
of the extra generation can be realized.

Our study has immediate experimental relevance. The
requirement that all couplings remain perturbative places
restrictive upper bounds on the $t'$, $b'$, $\nu'$ and $\tau'$ masses.
These are such that the question of an MSSM fourth generation may
be settled in the next few years by searches at
the Tevatron for an appropriate set of new signatures (\ie\ besides
$t\rightarrow b+W$). And certainly LEP-II will allow detection
of direct $\tau'+\bar \tau'$ and/or $\nu'+\bar \nu'$ production
for masses up to the maximum allowed.

\bigskip
\leftline{{\bf 2. The Renormalization Group Equations and Procedure}
\foot{We assume that the third and fourth generations are unmixed, but
later comment briefly on the (small) effects of allowing arbitrary mixing.}}
\smallskip

For a sequential fourth generation model
with the addition of a gauge singlet $\nu_R$ the one-loop
predictions for $M_U$ and $\alpha_3(\mz)$ are independent of the
number of generations, and there is only a mild dependence
of $M_U$ and $\alpha_3(\mz)$ on $N_g$ at two-loops, implying that
the gauge-coupling sector alone will not restrict $N_g$.

In contrast, $N_g$ does affect Yukawa coupling unification at one loop.
In order to accommodate a massive $\nu'$
we introduce the superpotential Yukawa interaction $\delta {\cal W}\equiv
\lambda_{\nu'} \hat L'\hat H_2 \hat \nu'_R$ where
$\hat L'=(\hat \nu',\hat \tau')_L$ and $\hat H_2$ is
the Higgs doublet superfield whose neutral
scalar component gives mass to up-type quarks and (now) the $\nu'$.
The RG equations obtained in the absence of the $\nu'$
\Ref\INOUEetal{K. Inoue, A. Kabuto, H. Komatsu, and S. Takeshita,
\ptpj{67}, 1889 (1982); Bjorkman and Jones, Ref. [9].}\ are modified
by: i) the RGE for the fourth generation neutrino
$y_{\nu'}$; and ii) the additional
$y_{\nu'}$ terms which appear in the RGE's for $y_t$ and $y_{t'}$ (via
Higgs field renormalization) and for $y_{\tau'}$ (via
fermion field renormalization) --- here, we have defined
$y_f\equiv \lambda_f^2/4\pi$.
More explicitly, for the third and fourth generation fermions we obtain
(at one-loop):
$${dy_{i}\over dt} = (A_{ji}y_{j}-\alpha_k B_{ki}){y_i}~,
\quad i={\rm fixed}\,,\eqno(1)$$
\noindent
where $i,j=(t,b,\tau,t',b',\nu'$,$\tau')=(1,2,3,4,5,6,7$),
$(\alpha_k)=(\alpha_1,\alpha_2,\alpha_3)$
(where $\alpha_i\equiv g_i^2/4\pi$), $t\equiv\ln (\mu)/2\pi$
and the $A_{ji},B_{ki}$ matrices are
$$
A=\pmatrix{6&1&0&3&0&3&0\cr 1&6&3&0&3&0&3\cr 0&1&4&0&1&0&1\cr
3&0&0&6&1&3&0\cr 0&3&3&1&6&0&3\cr 1&0&0&1&0&4&1\cr 0&1&1&0&1&1&4\cr},
\qquad B=\pmatrix{{13\over
15}&{7\over 15}&{9\over 5}&{13\over 15}&{7\over 15}&{3\over 5}&{9\over 5}\cr
3&3&3&3&3&3&3\cr {16\over 3}&{16\over 3}&0&{16\over 3}&{16\over 3}&0&0\cr}~.
\eqno(2)$$

Since $\lambda_{u,d}=\sqrt{2}m_{u,d}/\left[v(\sin\beta,\cos\beta)\right]$
(with $v=246\gev$, $\tanb\equiv v_2/v_1$), and
fourth generation masses must satisfy the LEP bounds
$m_{u,d}>\mz/2$, {\it all} of the fourth generation Yukawa couplings
must be relatively large at the weak scale:
$\lambda_i\gsim 0.25$. As a result, the Yukawas only remain perturbative
during evolution up to $M_U$ for a severely limited
range of $\tanb,\mt,m_{t'},m_{b'},m_{\nu'}$ and $m_{\tau'}$. Indeed,
we find that $\tanb$ must be $\lsim 3$ in order to avoid
singular Yukawa coupling behavior; large $\tanb$ solutions
are not possible when a fourth generation is included.
When the values of $\mt,m_{t'},m_{b'},m_{\nu'},m_{\tau'}$ are comparable, the
$y_{\tau'}$ and $y_{\nu'}$ couplings in Eq.~1 are the most
susceptible to singular ``Landau-pole'' behavior
since there is no large, negative gluon ($\alpha_3$)
term in their $\beta$-functions to control the growth
driven by the large, positive Yukawa terms. Only when $\mt$ is large
and $m_{\nu',\tau'}$ are near the LEP lower bound of $\sim\mz/2$
(see Sec. 4) do other Yukawas first go non-perturbative.  Then,
the (large) $t,t',b'$ Yukawas feed
one another's growth, and all receive additional boosts
from the $\nu'$ and $\tau'$ Yukawas, with the result that
$\lambda_t$, $\lambda_{b'}$ or $\lambda_{t'}$
can go non-perturbative sooner
than $\lambda_{\tau',\nu'}$ in evolving up to $M_U$.

Although the fourth generation Yukawa factor $3y_{b'}+y_{\tau'}$ appears in the
$\beta$-functions for $y_b$ and $y_\tau$ in Eq.~1, this common
fourth-generation down-sector (Higgs field renormalization) term cancels in
${d\over {dt}}({y_b\over {y_\tau}})$ yielding a result
identical to the three generation case.
Nonetheless, the presence of a fourth generation tends to
drive $m_b$ to unacceptably high values if
$R_{b/\tau}=1$ is required.\refmark{\Nanojones}\ To see why, first recall
that quark to lepton Yukawa ratios that are of
${\cal O}(1)$ at $M_U$ evolve to values that are larger than one
as one runs down to the weak scale (and below) due to the negative
$\alpha_3$ terms that are present in the quark Yukawa RGE's but
absent from lepton Yukawa RGE's. This effect is substantially
magnified for $N_g=4$ compared to $N_g=3$ because
gauge unification yields larger $\alpha_U$, and hence $\alpha_3(t<M_U)$,
when $N_g=4$. The indirect influence of
$y_{t'}$ and $y_{\nu'}$ on $y_t$ also affects
the low-energy $m_b/m_\tau$ value predicted for a given choice of
$\lambda_b/\lambda_\tau$ at $M_U$, but cannot compensate for
the $\alpha_3$ magnification effect.
Evolution of $R_{b'/\tau'}$ ($\equiv \lambda_{b'}(M_U)/\lambda_{\tau'}(M_U)$)
involves the fourth generation couplings
directly, and it is possible to have $R_{b'/\tau'}=1$ for sufficiently small
$\tanb$. In any case, as motivated in the introduction, we shall not require
$b/\tau$ and/or $b'/\tau'$ unification.

\bigskip
\leftline{\bf 3. Two-Loop RG Equation Results}
\smallskip

At two-loop order, the RG equations in both the gauge and Yukawa sectors
are fully coupled. For this study, two-loop Yukawa RG equations are used,
but in the gauge sector only the two-loop mutual gauge couplings are retained.
(The two-loop Yukawa effects on the gauge sector
are small and will be commented upon shortly.)
Further, we assume that the SUSY mass scale
and $\mz$ are the same, so that there is no intermediate decoupling
of SUSY partner fields. Gauge-coupling unification along with
the experimental central values of $\sin^2\theta_W(\mz)=0.2324$
\Ref\sinsq{See P. Langacker and N. Polonsky, Ref.~[\threshold].}\
and $\alpha_{QED}(\mz)=1/127.9$
then determines $\alpha_3(\mz)$; the result is $\alpha_3(\mz)=0.130$.
We also obtain $M_U=5.46\times 10^{16}\gev$ and
$\alpha_U=0.0936$.\refmark\sinsq\ The Yukawa equations are
then solved iteratively subject to requiring definite input values of $\tanb$
and $\mt(\mt)$, $m_b(m_b)$, $m_{\tau'}(m_{\tau'})$,
and $m_{\nu'}(m_{\nu'})$.\foot{Unless otherwise noted,
all masses referred to in the following are the running mass
$m(m)$.} A bound of $\lambda_f<3$ is imposed in order to
guarantee perturbative validity for the RGE's as defined by
a modest ratio of
two-loop to one-loop effects.\Ref\BARGERp{V. Barger, M. S. Berger and
P. Ohmann, Ref. [5].}\ The output comprises values of $m_{b'}(m_{b'})$
and $m_{t'}(m_{t'})$ that are consistent with the RGE's and the
perturbative bounds. An allowed choice for the $t'$ and $b'$ running
masses yields definite values for $R_{b/\tau}$ and $R_{b'/\tau'}$.

As discussed in Sec. 2, $\lambda_{\nu'}$ and
$\lambda_{\tau'}$ are intrinsically the most susceptible to
Landau-pole divergence in evolving up to $M_U$.
Indeed, $\lamnup$ or $\lamtaup$ are typically
driven to non-perturbative values unless
$m_{\nu'},m_{\tau'}$ are near
their lower bound of $\sim \mz/2$.  Thus, to explore large
$t,t',b'$ Yukawas and large $m_{b',t'}$ values, we employ large $\mt$, take
$m_{\tau'}=m_{\nu'}=50\gev$, and examine the region of perturbatively allowed
solutions in the $m_{b'}-m_{t'}$ plane for various values of $\tanb$.

\TABLE\yukawa{}

\FIG\figurei{
The allowed region in $m_{b'}$--$m_{t'}$ parameter space
is illustrated for $m_{\tau'}=m_{\nu'}=50\gev$, $m_b=4.6\gev$, and
$\mt=160\gev$. (All masses are $m(m)$ running masses.)
The perturbatively disallowed region is dotted. Within the allowed
undotted region, the horizontal solid (vertical dashed) lines correspond to
fixed values of $R_{b/\tau}$ ($R_{b'/\tau'}$) as labelled.
Results for $\tanb=1.5$ and $2.2$ are shown. The dash-dot lines
in the $\tanb=2.2$ window are contours of $\Delta\rho=0.002$, $0.004$ and
$0.006$ (in order of decreasing restrictiveness).
The two-loop Yukawa corrections discussed for Table~\yukawa\
were evaluated at the points marked by the $\times$'s.
}
\topinsert
\vbox{\phantom{0}\vskip 4.5in
\phantom{0}
\vskip .5in
\hskip -10pt
\special{ insert scr:figure1_4thgen.ps}
\vskip -1.45in }
\centerline{\vbox{\hsize=12.4cm
\Tenpoint
\baselineskip=12pt
\noindent
Figure~\figurei:
The allowed region in $m_{b'}$--$m_{t'}$ parameter space
is illustrated for $m_{\tau'}=m_{\nu'}=50\gev$, $m_b=4.6\gev$, and
$\mt=160\gev$. (All masses are $m(m)$ running masses.)
The perturbatively disallowed region is dotted. Within the allowed
undotted region, the horizontal solid (vertical dashed) lines correspond to
fixed values of $R_{b/\tau}$ ($R_{b'/\tau'}$) as labelled.
Results for $\tanb=1.5$ and $2.2$ are shown. The dash-dot lines
in the $\tanb=2.2$ window are contours of $\Delta\rho=0.002$, $0.004$ and
$0.006$ (in order of decreasing restrictiveness).
The two-loop Yukawa corrections discussed for Table~\yukawa\
were evaluated at the points marked by the $\times$'s.
}}
\endinsert

Fig.~\figurei\ illustrates the scope of the perturbatively-allowed solutions,
adopting $m_b=4.6\gev$\foot{Upper limits of $\mb(\mb)$ allowed
in Refs.~[\threshold,\JORGE] range from 4.45 to 4.9 GeV.}
and $\mt=160\gev$ (\ie\ $\mt(pole)=169\gev$).
\foot{Similar results apply for higher $\mt$.}
For $\tanb=1.5$ we find a nearly square allowed region extending
to $\mbp,\mtp \lsim 120-125\gev$, as shown in
Fig.~\figurei a. For $\tanb= 2.2$ (and higher\foot{We will shortly
establish an upper limit of $\tanb\lsim 3$.}), larger $\mtp$ values
$\lsim \mt$ are possible but only for $\mbp<\mz$, as
illustrated in Fig.~\figurei b.
For both $\tanb$ cases, $R_{b/\tau},R_{b'/\tau'}$ are simultaneously
near-maximal (but $<1$) in the large-$\mbp$, large-$\mtp$ corner of
parameter space. In Fig.~\figurei a, $m_{t'}^{max}$ at fixed $\mbp$
($m_{b'}^{max}$ at fixed $\mtp$) occurs when
$\lambda_t$ ($\lambda_{b'}$, with $\lambda_{\tau'}$ close behind)
goes non-perturbative.
In Fig.~\figurei b, $m_{t'}^{max}$
occurs when $\lambda_{t'}$ (followed very closely by $\lambda_{\nu'}$
and $\lambda_t$) goes non-perturbative; at $m_{b'}^{max}$
it is $\lambda_{\tau'}$ that goes non-perturbative.
For higher $m_{\nu'}=m_{\tau'}$ (\eg\ $\gsim 55\gev$ at $\tanb=1.5$)
it is $\lambda_{\tau'}$ that goes non-perturbative at all boundary points.

The heavy neutrino effects, included here for the first time in
this type of study, typically lower the $m_{t'}$ upper limit by about $5\gev$
for given input values for $\mt$, $\tanb$ and $m_{\tau'}$. For
specific allowed values of $m_{t'}$ and $m_{b'}$
they cause the value of $R_{b/\tau}$ ($R_{b'/\tau'}$) to rise by
$3-5\%$ ($20-30\%$) compared to the
massless neutrino case. Thus, the $\nu'$ couplings do not
change the qualitative
picture, but they are certainly not negligible.\foot{We have solved the
one-loop RG equations for the case of arbitrary $3-4$ generation
mixing, and find effects on the boundaries and the values of
$R_{b/\tau}$ at only the few percent level.}

\midinsert
 \titlestyle{\tenpoint
 Table \yukawa: Corrections to $\alpha_3(\mz)$, $\alpha_U$, $M_U$,
$R_{b/\tau}$ and $R_{b'/\tau'}$ due
to two-loop Yukawa effects in the gauge sector for the selected points
shown by the $\times$'s in Fig.~\figurei\ (with
$m_{\nu',\tau'}=50 \gev$ and $\mt=160 \gev$). The 0 superscript indicates
values before including two-loop effects.}
 \smallskip
\def\tstrut{\vrule height 12pt depth 4pt width 0pt}
 \thicksize=0pt
 \hrule \vskip .04in \hrule
 \begintable
$(m_{t'},m_{b'})$ (GeV) |   $\alpha_3/\alpha_3^0$ |   $\alpha_U/\alpha_U^0$ |
  $M_U/M_U^0$ |   $R_{b/\tau}/R_{b/\tau}^0$ |
 $R_{b'/\tau'}/R_{b'/\tau'}^0$ \cr
(95,95) (GeV) |  0.97 | 0.95 | 0.91 | 1.08 | 1.05 \cr
(140,70) (GeV) | 0.96 | 0.94 | 0.91 | 1.11 | 1.06
\endtable
 \hrule \vskip .04in \hrule
\endinsert

Table~\yukawa\ indicates the effects of
including two-loop Yukawa terms in the gauge-coupling RGE's at
the points marked by $\times$'s in Fig.~\figurei.
Since $\alpha_3$ is decreased,
allowed solution regions are slightly reduced and
$\rbtau$ and $\rbtaup$ are increased by up to 10\%.

\FIG\figureii{
In a), we display the maximal allowed values for $m_{t'},m_{b'}$ as a
function of $\tanb$ taking $m_b(m_b)=4.6\gev$, $\mt(\mt)=160\gev$
(corresponding to $\mt(pole)=169\gev$), and $\mtaup=\mnup=50$ or $75\gev$
(as indicated by the numbers in parenthesis):
the solid (dotted) curves correspond to
$m_{t'}^{max}$ ($m_{b'}^{max}$) assuming $\mbp=45\gev$ ($\mtp=45\gev$).
In b), we take $\mtaup=\mnup=45\gev$. The solid (densely dotted) curves give
$m_{t'}^{max}$ ($m_{b'}^{max}$) as a function of $\mt(\mt)$; the
dashed (sparsely dotted) curves show (on the right-hand $y$-axis)
the value of $\tanb$  at which
these respective maximal values are attained.
}
\topinsert
\vbox{\phantom{0}\vskip 4.5in
\phantom{0}
\vskip .5in
\hskip -10pt
\special{ insert scr:figure2_4thgen.ps}
\vskip -1.45in }
\centerline{\vbox{\hsize=12.4cm
\Tenpoint
\baselineskip=12pt
\noindent
Figure~\figureii:
In a), we display the maximal allowed values for $m_{t'},m_{b'}$ as a
function of $\tanb$ taking $m_b(m_b)=4.6\gev$, $\mt(\mt)=160\gev$
(corresponding to $\mt(pole)=169\gev$), and $\mtaup=\mnup=50$ or $75\gev$
(as indicated by the numbers in parenthesis):
the solid (dotted) curves correspond to
$m_{t'}^{max}$ ($m_{b'}^{max}$) assuming $\mbp=45\gev$ ($\mtp=45\gev$).
In b), we take $\mtaup=\mnup=45\gev$. The solid (densely dotted) curves give
$m_{t'}^{max}$ ($m_{b'}^{max}$) as a function of $\mt(\mt)$; the
dashed (sparsely dotted) curves show (on the right-hand $y$-axis)
the value of $\tanb$  at which
these respective maximal values are attained.
}}
\endinsert

Figure~\figureii a gives the maximum allowed $m_{t'}$ ($m_{b'}$) values
as a function of $\tanb$,
for $m_b(m_b)=4.6\gev$ and  $\mt(\mt)=160\gev$; these are attained by
setting $\mbp=45\gev$ ($\mtp=45\gev$), respectively. Results are shown
for the two choices $m_{\tau'}=m_{\nu'}=50$ and $75\gev$. Note that
$m_{b'}^{max}$ is attained just above the lower $\tanb$ limit
where $m_{t'}^{max}$ is forced below 45 GeV, while $m_{t'}^{max}$
occurs near the upper $\tanb$ end-point where $m_{b'}^{max}$ is forced
below 45 GeV.
For $m_{\tau'},m_{\nu'}=50\gev$ (as in Fig.~\figurei), Fig.~\figureii a
shows that non-perturbative values for all the $\lambda_i$ can only be
avoided for $1\lsim\tanb\lsim 3$. The LEP bounds
on the fourth generation are crucial in arriving at this
very restrictive limit of $\tanb\lsim 3$.
(The 3-generation-like $\tanb$
range, $\tanb\lsim 60$, is only attained in the phenomenologically disallowed
limit where all fourth-generation masses ---and, hence, Yukawas --- can
be simultaneously near zero, leaving only the
$N_g=4$ effect in the gauge RG equations.)
As $m_{\tau'},m_{\nu'}$ increase, the allowed $\tanb$
domain collapses further and $\mtp^{max}$ and $\mbp^{max}$ quickly
fall towards $\mz/2$, as illustrated by the $m_{\tau'}=m_{\nu'}=75\gev$
curves in Fig.~\figureii a.
At $\mt=160\gev$, $\mtaup=\mnup>83\gev$ is excluded
if we require $\mtp,\mbp\geq 45\gev$.

Of course, the allowed solution regions vary with $\mt$.
The absolute maximal values of $\mtp$ and $\mbp$ consistent with LEP limits
(found by setting $\mtaup,\mnup=45\gev$ and scanning in $\tanb$
as in Fig.~\figureii a), and the $\tanb$ values at which they are attained,
are plotted in Fig.~\figureii b as a function of $\mt$. We see that the highest
value of $\mt(\mt)$ for which there is a consistent perturbative
solution to the gauge-coupling and Yukawa RGE's is $\mt\sim 207\gev$.
At this $\mt$, $\mbp^{max}$ and $\mtp^{max}$ are simultaneously forced to the
imposed 45 GeV LEP limit and the $\tanb$ values at which these maxima are
achieved converge to a value just above 3.
As $\mt$ is decreased, $\mbp^{max}$ and $\mtp^{max}$ both rise rapidly
and the $\tanb$ value at which $\mbp^{max}$ is attained falls rapidly
relative to the rather constant $\tanb\sim 3$ value at which $\mtp^{max}$
is attained. Note that $\mtp>\mt$ becomes possible for $\mt\lsim 170\gev$.
In analogous fashion, we determine the maximum allowed $\mtaup=\mnup$
value subject to requiring $\mtp,\mbp>45\gev$. The results
for $\mt=130$, $150$, $170$, $190$, $207\gev$ are
$\mtaup=\mnup =90.7$, $86.1$, $79.0$, $67.9$, $45.0\gev$, respectively.
These maxima are achieved at $\tanb=1.26$, $1.37$, $1.56$, $1.94$
and $3.02$, respectively.

Finally, let us discuss whether or not some or all of the Yukawa
couplings are naturally near their fixed-point (defined by
zero derivative) at scale $\mz$.
In the case of three generations, requiring $\rbtau=1$
(and $\mb(\mb)=4.25\gev$) implied that
$\mt$ was close to its ($\tanb$-dependent) fixed-point value at $\mz$,
and evolved to large values at $M_U$.\refmark{\fixedpoint,\BARGERp}\
This connection between fixed-point behavior and Yukawa unification
is modified for four generations. We have already noted that
$\rbtau<1$ for all perturbatively allowed $\mt,\mtp,\mbp$
choices. Nonetheless, for small enough $\tanb$, $\mnup$,
and $\mtaup$ choices, $\mt,\mtp,\mbp$ values within the perturbative domain
can be found such that $\lamt,\lamtp,\lambp$
are all simultaneously very close to their fixed-point
values.\foot{$\lamtaup$ and $\lamnup$ cannot be at a fixed
point at scale $\mz$ since all fourth-generation Yukawas must be large
(due to LEP and other lower bounds on the associated masses)
and since there is no large $\alpha_3$ term to cancel the large
Yukawa coupling terms in the $y_{\tau'}$ and $y_{\nu'}$ evolution equations.}
For $\mnup=\mtaup=50\gev$ and $\tanb=1.5$ the
$\lamt$, $\lamtp$ and $\lambp$ derivatives are simultaneously zero for
$\mt,\mtp,\mbp=162.2$, 126.3, $118\gev$.
\foot{In general, specifying $\tanb,\mtaup,\mnup$ yields three
equations in the three unknown $\lamt$, $\lamtp$ and $\lambp$
Yukawas; solving yields definite values for the low-scale $t,t',b'$
Yukawas and, hence, masses.}
The corner of the $\tanb=1.5,\mt=160\gev$
Fig.~\figurei a where $\mbp$, $\mtp$, $\rbtau$ and $\rbtaup$ are
all near maximal is very close to the $\tanb=1.5$ fixed-point solution.
For still smaller $\tanb$, one can achieve $\rbtaup\sim 1$ and somewhat
larger $\rbtau$. For example, for $\tanb=1.2$ and $\mt=160\gev$
the parameter-space-corner
point $\mtp,\mbp=78,136\gev$ yields $\rbtau,\rbtaup=0.65,1.1$.
Such points are even closer to a $\lamt,\lamtp,\lambp$ fixed point.
In contrast, for $\tanb\gsim 2$ it is not possible to achieve large
$\rbtaup$ or approach fixed-point behavior;
a small $\rbtaup$ value implies a $\lambp$ value well below 1
as $\lamtaup$ passes (as $\mbp$ is increased) the perturbative limit of 3.
In general,
adjusting $\mt,\mtp,\mbp$ so that both $\rbtau$ and $\rbtaup$ are
as large as perturbatively allowed at a given $\tanb$ implies that
$\lamt,\lamtp,\lambp$ will all be as close to $\mz$-scale fixed-point
values as perturbatively possible --- however, the fixed point
lies increasingly beyond the perturbative domain as $\tanb$ is increased
(which forces $\lamtaup$ to reach its perturbative limit of 3 at $\mbp^{max}$
while other Yukawas are still small).

\bigskip
\leftline{\bf 4. Experiments and a Fourth Generation}
\smallskip

The bound $m_f\gsim \mz/2$ on any new sequential fermion, quark or lepton,
that has been set at LEP\Ref\OPAL{OPAL Collaboration, M. Z. Akrawy et al.
\plbj{236}, 364 (1990); ALEPH Collaboration, D. Decamp et al., {\ibid},
{\bf 236}, 511 (1990).}\ is the firmest experimental constraint on the
fourth generation masses. A full analysis of
constraints on the masses of fourth family quarks with small
mixings to lower-mass families from precision electroweak observables
might prove useful.
\Ref\PDG{Particle Data Group, K. Hikasa et al., \prdj{45}, 51
(1992), vol. 11, pp. 159-163.}\
\ \foot{Contributions from spin-0 superpartners are typically much smaller.}
Indeed, requiring $\Delta\rho (t'-b')\lsim 0.002$
\foot{The $\tau'$ and $\nu'$ are close to the
lower bound $\mz/2$ and are ignorable for our purposes.}
(\ie\ $\Delta T\lsim 0.25$),
would eliminate the large-$m_{t'}$ small-$m_{b'}$
corner of the $\mt=160\gev$ parameter space ---
see the contour in Fig.~\figurei b. Also, a roughly
degenerate fourth family with masses $\gsim \mz$ yields $\Delta S\sim 2/3\pi$,
a shift comparable to that obtained by changing the Higgs mass from 100 GeV
to 1 TeV in the SM. However, because the MSSM RGE
perturbative limits imply (correlated) upper bounds on
the $b'$ and $t'$ masses, direct experimental constraints from the Tevatron
can be much more significant, depending upon the
relative size of the $t'$ and $b'$ masses.
\foot{In the following experimental constraint discussion, the masses
referred to should be thought of as pole masses.}

We consider first the case of $m_{t'}>m_{b'}$.
Constraints on the $b'$ depend greatly upon the manner in which it decays.
If the $b'$ has significant mixing with the 2nd or 1st generation
then $b'\rta c,u +W$ (where the $W$ may be real or virtual)
decays will be dominant.
The published CDF collaboration data\Ref\cdfbprime{F. Abe \etal,
\prlj{64}, 147 (1990), \prlj{68}, 447 (1992). T. Trippe, Particle Data Group,
private communication.}\
limits the mass of a $b'$ that decays primarily to a $c$ ($u$) quark:
$m_{b'}>85\gev$ at $95\%$ C.L., using
the same $e\mu+X$ final state search that excluded the $t$ in this mass range.
Their latest data\Ref\newtopdata{F. Abe \etal, Fermilab-PUB-94/097-E.}
in the dilepton channel probably can be used to increase this limit.
However, if unmixed with lower generations, the $b'$ will decay via
flavor-changing neutral current (FCNC) channels:
$b'\rightarrow b \gamma,b g$ or $bZ^*$ for $m_{b'}<\mz$, with $b'\rta bZ$
becoming dominant for $m_{b'}>\mz$.\foot{We temporarily ignore
the decays $b'\rightarrow h b$
(where $h$ is the light MSSM Higgs boson) that would dominate if allowed.
\Ref\BpHb{W. -S. Hou and R. Stuart, \prdj{43}, 3669 (1991).}}
These decays yield distinctly different
signatures than the lepton-plus-jet signatures of charged current decays
that are already mostly excluded. New search strategies are required.
\Ref\novelb{See Agrawal and Hou, and Agrawal, Ellis and Hou, and references
therein, Ref.~[\AGRAWALetal].}\ For example, when $m_{b'}<\mz$ the process
$p\bar p\rightarrow b'\bar b'\rightarrow b\gamma+\bar b\gamma$ yields
a significant rate for $2{\rm jet}+2\gamma$ final states.
Also present, and at a higher rate such that
one might be able to afford to $b$-tag,
would be $\gamma+3{\rm jet}$ final states. Present CDF data
in combination with the data from the ongoing run may
combine to set rather firm $b'$ limits.\Ref\HOUSTON{J. Houston and R. Blair,
private communication.}\
If $m_{b'}>\mz$, but below our upper limits of roughly $140-150\gev$,
then $b'\bar b'\rta b \bar b ZZ$ yields a significant excess
in the rate for real $Z$ production.
For example, if $m_{b'}=130\gev$ and $BR(b'\rta b Z)\sim 0.8$,
there are roughly 60 $b'\bar b'$
events for $L=19.3\pbi$ in which at least one of the $Z$'s decays
to $e^+e^-$ or $\mu^+\mu^-$. Interestingly, in Ref.~[\newtopdata] it
is stated that there is a slight excess of $Z$'s (in association
with a tagged $b$) that is not inconsistent
with production and FCNC decay of a $b'$ with $m_{b'}\sim 150\gev$.

\TABLE\cdfrates{}
\midinsert
 \titlestyle{\tenpoint
 Table \cdfrates: For $L=19.3\pbi$, we tabulate the number of $t'\overline{t'}$
events passing our approximations to the
CDF top-quark discovery cuts in the single lepton+$b$-tag and dilepton modes.}
 \smallskip
\def\tstrut{\vrule height 12pt depth 4pt width 0pt}
 \thicksize=0pt
 \hrule \vskip .04in \hrule
 \begintable
\ | \multispan{4} \tstrut\hfil Lepton + $b$-tag Mode \hfil |
\multispan{4} \tstrut\hfil Dilepton Mode \hfil \cr
\ | \multispan{4} \tstrut\hfil $m_{b'}$ (GeV)  \hfil |
\multispan{4} \tstrut\hfil $m_{b'}$ (GeV) \hfil \nr
$m_{t'}$ (GeV) | 50 & 80 & 110 & 130 | 50 & 80 & 110 & 130 \cr
  160 |  2.8 & 2.2 & 0.5 & 0.2 | 0.7 & 0.5 & 0.06 & 0.004 \nr
  130 |  7.3 & 2.8 & 0.07 & $-$ | 1.2 & 0.3 & 0.0006 & $-$ \nr
  100 |  7.9 & 0.5 & $-$ & $-$ | 0.8 & 0.003 & $-$ & $-$
\endtable
 \hrule \vskip .04in \hrule
\endinsert

Consider next the $t'$. If $m_{t'}$ is significantly larger than $m_{b'}$,
\eg\ $m_{t'}\simeq \mt\gsim 150\gev$ and
$m_{b'}\lsim 75\gev$ (as in the corner region of Fig.~\figurei b),
then $m_{t'}$ lies above the published $D0$ limit on $\mt$,
\Ref\Dzero{D0 Collaboration, S. Abachi et al., \prlj{72}, 2139 (1994).}
and is in the region favored for $\mt$
by the recent CDF analysis.\refmark\newtopdata\
The corresponding $t'$ production and decay process
is especially interesting, giving a complex set of
possible final states with six and eight jets.
We have performed a rough Monte Carlo study of the
number of $t'\bar t'$ production and decay events in the
dilepton-plus-jets and lepton-plus-jets channels
(keeping only those final states that
do not contain any extra leptons or photons from the $b'$ decays),
using the quoted CDF luminosity and approximating their cuts and $b$-tagging
procedures. Branching ratios of $b'$ decays are estimated from
Ref.~[\novelb].
A summary of the predicted event rates appears in Table~\cdfrates.
\foot{For comparison, at $\mt=170\gev$, we obtain an uncut $t\anti t$
cross section of 4 pb. Of the $\sim 80$ events predicted for $L=19.2\pbi$,
roughly 4 events pass our single-lepton plus $b$-tag cuts, and 0.8 events pass
our dilepton cuts.}
The rates are not
inconsistent with the most recent CDF data, especially given that
the latter show a somewhat larger
number of events than expected for the favored mass,
$\mt\simeq 170\gev$.\refmark\newtopdata\
However, the $t'$ events do not necessarily provide a natural explanation
for the excess; they contain significantly more jets and yield (following
the CDF procedure) significantly lower reconstructed `top' mass values.
A thorough ``second top" study by the CDF group would be most interesting.

The corner region of Fig.~\figurei a, where
$R_{b/\tau}$ and $R_{b'/\tau'}$ are nearest to 1,
contains the case $m_{t'}\simeq m_{b'}\simeq 110-120\gev$.
Again, if $b'\rightarrow c,u$ and
$t'\rightarrow b,s,d$ are adequately suppressed, then for $m_{t'}>m_{b'}$
the $t'$ decays to $b'$
along with a rather low mass system of $W$-decay products.
The $b'$ would then decay preferentially to
$Z+$light quark, giving a dilepton signal that would probably have
been detected.

The above discussions would be altered
should the light, neutral Higgs boson ($h$) of the MSSM have mass less than
$m_{b'}$, since then $b'\rightarrow h b$ decays would certainly (most probably)
be dominant for $\mbp$ smaller (greater) than $\mz+\mb$.\refmark\BpHb\
The possibility of losing signals
such as the above-noted dilepton channel cannot be discounted.
Although at tree-level $m_h$
is small at the small $\tanb$ values allowed in the presence of a
fourth generation, the latter can add substantially to the radiative
corrections from the top/stop sector (depending upon superpartner masses).

{}From Figs.~\figurei\ and \figureii\ we see that
the RG equations also have solutions
with $m_{t'}<m_{b'}<\mt$. Experimental constraints in this case depend on
how the $t'$ decays. We regard it as quite likely that 3rd--4th
generation mixing will be large enough that the $t'\rightarrow bW$ decays
will be dominant. In this case, a $t'$ with $m_{t'}<130\gev$
(which limit applies for all $m_{t'}<m_{b'}$ regions allowed by the MSSM
RGE's) has been excluded by the normal $t$-quark searches. However, constraints
could be much weaker if 3rd--4th generation mixing is extremely small
and the $t'$ decays were of the ($t'\rta cX$) FCNC variety.
\REF\progress{Work in progress.}
For further details, see Ref.~[\progress].

\bigskip
\leftline{\bf 5. Conclusions.}
\smallskip

In summary, our analysis has demonstrated
that there is still room to add a fourth generation in the gauge-unified MSSM
without violating perturbative or experimental constraints.
However, the range of solutions
is restricted even when $\lambda_b/\lambda_\tau$ and
$\lambda_{b'}/\lambda_{\tau'}$ are unconstrained at $M_U$.
In terms of running masses, if $\mt>150\gev$
the fourth generation $\tau'$ and $\nu'$ masses cannot lie
above $86\gev$, and they can only be this large if
$\mtp,\mbp$ are near the the LEP lower bound of $\mz/2$.
For the quarks, $m_{t'}\lsim \mt$ if $\mt\gsim
170\gev$ --- maximal $\mtp$ is attained for minimal
$\mnup,\mtaup$, \ie\ $\sim \mz/2$. Solutions with
$m_{t'}\sim \mt\sim 160-170\gev$ are possible for $\tanb\sim 2.2$; they
require $m_{b'}<\mz$. Although
our Monte Carlo study shows that these mass values could
give rise to extra ``top-like events", there is no immediate conflict
with the current CDF top signal analysis. However,
more data will greatly increase the constraints.
For $\mt\sim 160-170\gev$, $\tanb\sim 1.5$ yields
solutions with $m_{t'}\gsim m_{b'} \simeq 110-120\gev$.
Such a case could be hidden in the data
if $m_{t'}-m_{b'}$ is small enough that the $W^*$ decay products
are very soft and if $m_h<\mbp-\mb$ so
that the $b'$ decays mainly to Higgs boson plus $b$ quark.
This latter case is also an example where $\lambda_{t,t',b'}$ fixed-point
behavior comes close to being achieved, and
$\lamb/\lamtau,\lambp/\lamtaup$ are large (choosing the relatively
large value of $\mb(\mb)=4.6\gev$).
For still smaller $\tanb$ ($\sim 1.2$), one can approach
even nearer a $\lambda_{t,t',b'}$ fixed point and obtain
$\lamb/\lamtau\gsim 0.65$ and $\lambp/\lamtaup\sim 1$. However,
the corresponding $\mbp,\mtp$ values are
typically such that $\mbp>\mtp$ and $\mtp<\mz$, a scenario that would have
escaped detection only if the $t'\rta b W^{(*)}$ decays are unexpectedly
suppressed.

To ``tighten the noose'' on the fourth generation, a variety of theoretical and
experimental questions should be pursued. For example,
how is electroweak symmetry breaking affected, and how do detailed fits
of precision electroweak data change when $N_g=4$?
We believe that our study has shown that these questions
are of considerable theoretical and experimental interest, and should be
explored further.

\bigskip

\centerline{\bf Acknowledgments}
\smallskip
We wish to thank R. Blair and J. Huston for communicating the status of
their $b'$ analysis to us. DWM thanks the U.C. Davis Department of Physics
for its hospitality during his '93-'94 sabbatical.
This work has been supported in part by
Department of Energy grants \#DE-FG03-91ER40674, \#DE-FG02-85ER40214,
and by Texas National Research Laboratory grant \#RGFY93-330.

\bigskip

\refout
\bye